\documentclass[12pt]{reportj}
\usepackage{deluxetablej}
\usepackage{hyperref} 
\usepackage{amsmath}
\usepackage{placeins} 
\makeatletter
\def\@to{to}
\makeatother
\usepackage{times}
\usepackage{graphicx}
\usepackage{xcolor}
\usepackage{float}
\usepackage{tocloft}
\usepackage{fancyheadings}
\emergencystretch=\maxdimen
\usepackage{datetime}
\newdateformat{ddmonthyyyy}{\THEDAY\ \monthname[\THEMONTH]\ \THEYEAR}

\renewcommand\thesection{\arabic{section}}
\renewcommand\thesubsection{\thesection.\arabic{subsection}}

\def\ssection#1{\setcounter{subsection}{0} \refstepcounter{section} \section*{\hbox to \hsize{\large\bf \arabic{section}. #1\hfill }}\label{sec} \addcontentsline{toc}{section}{\arabic{section}. #1}}
\def\ssubsection#1{\setcounter{subsubsection}{0} \refstepcounter{subsection}\subsection*{\hbox to \hsize{\normalsize\bfseries\itshape \arabic{section}.\arabic{subsection} #1\hfill}}\label{subsec} \addcontentsline{toc}{subsection}{\arabic{section}.\arabic{subsection} #1}}
\def\ssubsubsection#1{\refstepcounter{subsubsection}\subsection*{\hbox to \hsize{\normalsize\it \arabic{section}.\arabic{subsection}.\arabic{subsubsection} #1\hfill}}\label{subsubsec} \addcontentsline{toc}{subsubsection}{\arabic{section}.\arabic{subsection}.\arabic{subsubsection} #1}}

\def\ssectionstar#1{\section*{\hbox to \hsize{\large\bf #1\hfill}} \addcontentsline{toc}{section}{#1}}
\def\ssubsectionstar#1{\subsection*{\hbox to \hsize{\normalsize\bfseries\itshape #1\hfill}} \addcontentsline{toc}{subsection}{#1}}
\def\ssubsubsectionstar#1{\subsection*{\hbox to \hsize{\normalsize\it  #1\hfill}} \addcontentsline{toc}{subsection}{#1}}

\renewcommand{\cftaftertoctitle}{%
\mbox{}\hfill{\normalfont Page}}
\begin{document}

~\\

\vspace{-2.4cm}
\noindent\includegraphics*[width=0.295\linewidth]{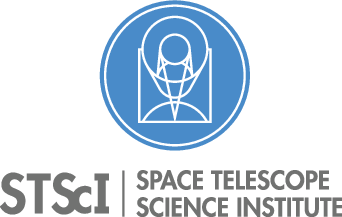}

\vspace{-0.4cm}

\begin{flushright}
    {\bf Instrument Science Report STIS 2026-06}
    
    \vspace{1.1cm}
    
    {\bf\Huge Updates to STIS NUV Dark Calibration}
    
    \rule{0.25\linewidth}{0.5pt}
    
    \vspace{0.5cm}
    
    Jacqueline Brown $^1$, Sean Lockwood $^1$, TalaWanda Monroe $^1$, Amy Jones $^1$, Joleen Carlberg $^1$
    \linebreak
    \newline
    \footnotesize{$^1$ Space Telescope Science Institute, Baltimore, MD\\}
    
    \vspace{0.5cm}
    
     25 September 2026
\end{flushright}

\vspace{0.1cm}

\noindent\rule{\linewidth}{1.0pt}
\noindent{\bf A{\footnotesize BSTRACT}}

{\it \noindent We present new superdark calibration files (DARKFILEs) and a new dark correction table (TDCTAB) for use with data taken by the Space Telescope Imaging Spectrograph near-ultraviolet microchannel array (STIS NUV-MAMA) detector for HST Cycle 19 onward. These superdarks include 12 years of NUV dark exposures and make use of an updated temperature scaling equation developed in 2013 to more accurately characterize the dark rate. In tested datasets, the new superdarks provide an average 3\% increase in SNR and smooth out a ring artifact left previously by oversubtraction of the dark rate. The new superdarks and TDCTAB have been delivered to the Calibration Reference Data System (CRDS) and have been used to recalibrate all affected datasets in the Mikulski Archive for Space Telescopes (MAST).}

\vspace{-0.1cm}
\noindent\rule{\linewidth}{1.0pt}

\renewcommand{\cftaftertoctitle}{\thispagestyle{fancy}}
\tableofcontents


\newpage

\vspace{-0.3cm}
\ssection{Introduction}\label{sec:Introduction} 
The Space Telescope Imaging Spectrograph (STIS) near-ultraviolet (NUV) multi-anode microchannel array (MAMA) is a photon-counting detector. For the NUV-MAMA, the primary contributor to detector background counts is the dark current. Most of the dark current in the MAMAs is caused by phosphorescence of impurities in the MgF$_2$ detector
faceplate. The behavior of the dark current is affected by both short-term variability and long term variability. The short term variability is exponentially linked to detector temperature as the meta-stable states are thermally excited to an unstable level and can more easily decay to the ground state. The long-term variability is influenced by the thermal history of the detector and incident high-energy particle flux (\href{https://hst-docs.stsci.edu/stisdhb}{STIS DHB Ch 3.4 and 4.1, 2024}), including charged particle impacts from passages through the South-Atlantic Anomaly (SAA) (\href{https://www.stsci.edu/files/live/sites/www/files/home/hst/instrumentation/stis/documentation/instrument-science-reports/_documents/201103.pdf}{STIS ISR 2011-03}). The dark current for the NUV-MAMA detector is mostly stable over time (\href{https://www.stsci.edu/files/live/sites/www/files/home/hst/instrumentation/stis/documentation/instrument-science-reports/_documents/201103.pdf}{STIS ISR 2011-03}; \href{https://www.stsci.edu/files/live/sites/www/files/home/hst/instrumentation/stis/documentation/instrument-science-reports/_documents/2013_01.pdf}{STIS ISR 2013-01}), but temporary increases in the dark rate can occur when the detector has been turned off for an extended period (also known as a safing event). The largest spike in dark rate occurred after the longest STIS safing, which lasted from the failure of Side-2 electronics in 2004 until operational recovery following Servicing Mission 4 (SM4) in 2009.

The \textbf{calstis} data calibration pipeline subtracts dark current from NUV-MAMA images using ``superdarks" (DARKFILEs), created by combining individual dark exposures, that are scaled with a relation encoded in the NUV Dark Correction Table (TDCTAB). In the dark signal subtraction calibration step (DARKCORR), the average count rate is calculated for a given exposure using the equation specified in Section \ref{sec:TDCTAB} and then divided by the mean count rate in the dark reference image to give a scale factor. The scale factor is applied to the superdark before it is subtracted from the uncalibrated science data. 

\begin{figure}[H]
    \centering
    \includegraphics[width=\linewidth]{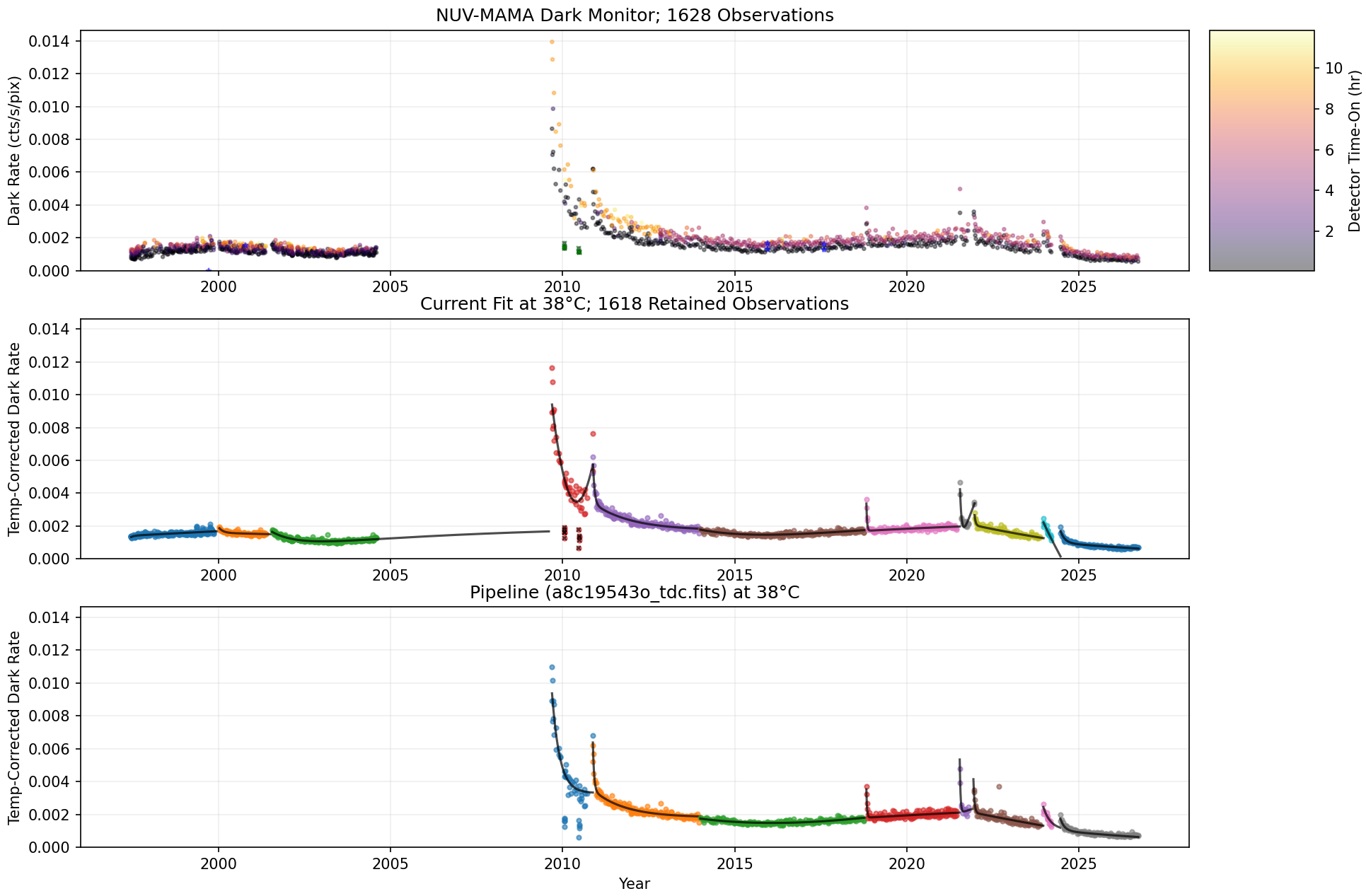}
    \caption{The NUV-MAMA dark monitor (an up-to-date monitor can be found at the \href{https://www.stsci.edu/hst/instrumentation/stis/performance/monitoring}{STIS Monitoring website}). The top plot contains the raw dark count rate for all STIS dark observations over time, including pre-SM4, and shows the number of hours the detector has been on before the dark observation. The middle plot contains the temperature-corrected dark count rate, with outliers removed, at an assumed nominal temperature of $38^{\circ}$ C, including pre-SM4, with estimated fits for each epoch between safings or significant changes in dark rate. The bottom plot contains the temperature-corrected dark count rate at an assumed nominal temperature of $38^{\circ}$ C, not including pre-SM4, with the current TDCTAB fits. Each colored segment in the bottom two plots represents a different fit.}
    \label{fig:dark_monitor}
\end{figure}

The first post-SM4 superdark was created from data taken in 2009 during the period of high dark current after operational recovery. Over time, the dark current declined to near pre-SM4 levels, between 0.001 and 0.002 cts/s/pix on average, and has been mostly stable outside of much shorter-term safing events (Figure \ref{fig:dark_monitor}).

The spatial structure of the NUV-MAMA is known to be stable (see Figure 4 in \href{https://www.stsci.edu/files/live/sites/www/files/home/hst/instrumentation/stis/documentation/instrument-science-reports/_documents/201103.pdf}{STIS ISR 2011-03}) and due to the small number of counts involved, the dark current is a minor contributor to overall error. This means the 2009 superdark continues to meet calibration specifications within the listed accuracies in the \href{https://hst-docs.stsci.edu/stisihb/chapter-16-accuracies/16-1-summary-of-accuracies}{IHB}, but there are some visible artifacts to be improved upon in more recent data that have minor impacts on signal-to-noise (SNR).

\FloatBarrier
\newpage

\lhead{}
\rhead{}
\cfoot{\rm {\hspace{-1.9cm} Instrument Science Report STIS 2026-06 Page \thepage}}

\vspace{-0.3cm}
\ssection{Data Collection: NUV MAMA Dark Monitoring Program}\label{sec:Data}
The STIS NUV-MAMA routine dark calibration program obtains two 1300 s\footnote{Before Cycle 20, the program included three orbits of paired 1314 s exposures to gather more information on the short-term temperature dependence.} darks every other week. The pairs of exposures are linked so that they are taken about six hours apart in the same SAA-free interval. This pairing of exposures separates long- and short-term temporal variability from temperature-dependent changes. Data from this program are used to monitor changes in the dark rate over time, produce updated values for the Exposure Time Calculator (ETC), and create reference files for calibration.

Measurements from 12 programs were used to create four new superdarks for the years 2011 to 2023 (Table \ref{tab:programs}). Anomalous datasets were not included in the creation of new superdarks. Anomalous data were identified by unusually short exposure times (less than a quarter of the planned exposure time of 1300 s), an exposure time of 0 s, or as a significant outlier in the NUV dark monitor (Figure \ref{fig:dark_monitor}). Only 16 files were excluded by these criteria.

Darks are obtained in TIME-TAG mode and the full exposure is converted to ACCUM before being used in the monitor or for creating reference files. TIME-TAG is high-resolution by default, meaning the raw darks (ACCUM-converted) are 2048x2048 pixels and in units of counts.

\begin{table}[htbp]
\centering
\caption{List of NUV Dark Calibration Programs}
\begin{tabular}{|l|l|l|l|l|}
\hline
\textbf{Cycle} & \textbf{Proposal} & \textbf{Dates of Observations} & \textbf{Number of Darks} & \textbf{Total Exposure Time (sec)} \\ \hline
19 & 12776 & 11/06/2011 - 10/16/2012 & 58 & 75484 \\ \hline
20 & 13147 & 11/04/2012 - 10/21/2013 & 58 & 75400 \\ \hline
21 & 13550 & 11/04/2013 - 10/20/2014 & 51 & 65000 \\ \hline
22 & 13996 & 11/08/2014 - 10/19/2015 & 52 & 67600 \\ \hline
23 & 14431 & 11/01/2015 - 10/17/2016 & 52 & 67600 \\ \hline
24 & 14835 & 10/30/2016 - 10/30/2017 & 54 & 70200 \\ \hline
25 & 14974 & 11/14/2017 - 11/04/2018 & 52 & 67338 \\ \hline
26 & 15563 & 11/16/2018 - 10/28/2019 & 52 & 67600 \\ \hline
27 & 15752 & 11/15/2019 - 10/26/2020 & 52 & 67600 \\ \hline
28 & 16354 & 11/09/2020 - 10/26/2021 & 46 & 59125 \\ \hline
29 & 16561 & 12/13/2021 - 10/29/2022 & 49 & 63700 \\ \hline
30 & 16962 & 11/10/2022 - 10/23/2023 & 52 & 67600 \\ \hline
\end{tabular}
\label{tab:programs}
\end{table}
\FloatBarrier

\vspace{-0.3cm}
\ssection{Updates to Characterizing NUV Dark Rate}\label{sec:TDCTAB}
The NUV-MAMA time and temperature dependence varies much more strongly than the spatial dependence. To model and correct for the time-and-temperature dependence, the NUV Dark Correction Table (TDCTAB) encodes an estimate of the scaling relation for the STIS NUV DARKFILEs as a function of detector temperature ({\tt OM2CAT}) and observation date (see \href{https://newcdbs.stsci.edu/doc/Section9.html#dark-correction-table-tdc-unique-name-tdc-fits}{HST Reference Files Information} for more details). The post-SM4 file uses the relation described in \href{https://hst-docs.stsci.edu/stisdhb/chapter-3-stis-calibration/3-4-descriptions-of-calibration-steps#id-3.4DescriptionsofCalibrationSteps-3.4.73.4.7DARKCORR:DarkSignalSubtraction}{STIS DHB Ch 3.4.7} and  \href{https://www.stsci.edu/files/live/sites/www/files/home/hst/instrumentation/stis/documentation/instrument-science-reports/_documents/2013_01.pdf}{STIS ISR 2013-01}:
\begin{equation} \label{eq:1}
\begin{split}
\mbox{count\_rate} = A1 \cdot \exp(-(D - DATE0) / D1) + A2 \cdot \exp(-(D - DATE0) / D2)\\
+ TA + TB \cdot (\mbox{\tt OM2CAT} - TEMP0) + TC \cdot (\mbox{\tt OM2CAT} - TEMP0)^{2}
\end{split}
\end{equation}
where $A1$ and $A2$ are the amplitude constants (count rate in cts/s/pixel) and $D1$ and $D2$ are the decay constants (in days) of the first and second exponential terms. $TA$, $TB$, and $TC$ are the coefficients of the quadratic temperature function (in count rate, count rate per degree, and count rate per degree-squared, respectively), which has been shifted relative to $TEMP0$. $TEMP0$ defines a common reference temperature of $38^{\circ}$ C. $DATE0$ defines the break points in MJD between piecewise components of the model, and $D$ is the date in MJD of the observation to be scaled by the equation. 

A a new post-SM4 TDCTAB, {\tt 92l21262o\_tdc.fits}, was delivered to CRDS on 2025-02-21.  It was derived from the previous version ({\tt 86715233o\_tdc.fits}).   Following monitoring observations of detector performance, we appended two new break points for MJDs 60297.29 and 60478.63, or 2023-12-19 and 2024-06-17, respectively. Prior fits remain unchanged.

\begin{figure}[H]
    \centering
    \includegraphics[width=\linewidth]{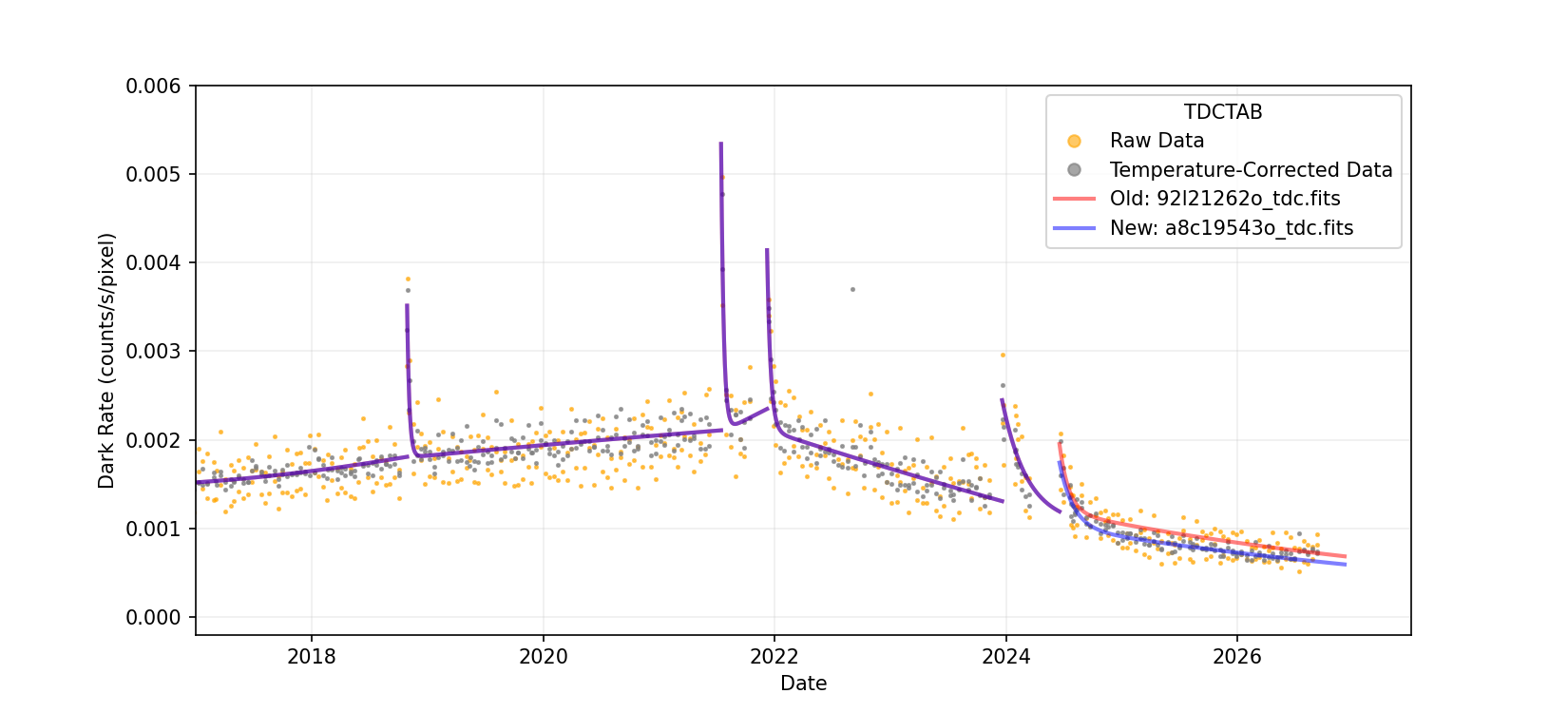}
    \caption{Recent NUV-MAMA dark rates (orange), temperature-corrected to $38^{\circ}$ C (black), and fit with dual-exponential fits (blue). The older fits (red) overlaps with the current fit (blue) in the last segment.}
    \label{fig:tdctab_fit}
\end{figure}

We fit a global linear temperature model (fixing $TC=0$) and used it to detrend data (\href{https://archive.stsci.edu/proposal_search.php?id=17391&mission=hst}{Jones, 2023}; \href{https://archive.stsci.edu/proposal_search.php?id=17646&mission=hst}{Brown, 2024}) in the time periods, and then fit a dual exponential model to each time window (see Figure \ref{fig:tdctab_fit}).  The newly corrected discontinuities in the dark rate resulted from instrument temperature swings during safing events.

The last segment starting at MJD 60478.63 (2024-06-17) was refit and delivered to CRDS on 2026-08-12. The new TDCTAB, {\tt a8c19543o\_tdc.fits}, was derived from the previous version, {\tt 92l21262o\_tdc.fits}. The temperature of the last segment was fit independently to constrain the baseline dark level.  

New DARKFILEs derived using the 2025-02-21 TDCTAB were delivered concurrently for reprocessing the post-SM4 STIS NUV-MAMA archive. While TDCTAB changes quantitatively affect only data taken after 2023-12-19, the corresponding DARKFILEs will affect data back to 2012-06-27.

\FloatBarrier

\vspace{-0.3cm}
\ssection{Reference File Creation}\label{sec:Creation}
\vspace{-0.3cm}
\ssubsection{File Production Steps} \label{subsec:production_steps}
The reference dark frame, or DARKFILE, contains a primary header and three data array extensions: the dark count rate (`SCI'), the statistical uncertainty on the dark rate (`ERR'), and the data quality flags (`DQ'). Dark data from Cycles 19 through 30 (Table \ref{tab:programs}) were used to create new DARKFILEs in the following production steps:\\

\noindent
Dark Count Rate:
\begin{itemize}
    \item Raw darks are first multiplied by a temperature scaling factor determined by the temperature correction equation with parameters derived from the TDCTAB (see Section \ref{sec:TDCTAB} for more information on the temperature correction equation). 
    \item Temperature-scaled darks are then summed pixel by pixel into a single 2048x2048 array and divided by total exposure time to be in cts/s.
    \item Summed data are then mean normalized to 2.84e-4 cts/s and saved to the first extension, `SCI'\footnote{The mean normalization level is consistent with DARKFILEs from 2002 onward (\href{https://www.stsci.edu/files/live/sites/www/files/home/hst/instrumentation/stis/documentation/instrument-science-reports/_documents/201103.pdf}{Zheng et al., 2011}).}.
\end{itemize}

\noindent
Error:
\begin{itemize}
    \item Error is given purely by counting statistics: $\textrm{err} = \sqrt{c}/t$, where $c$ is the total number of counts in a given pixel and $t$ is the total exposure time of the superdark. To calculate the error, counts are summed from the raw darks and are neither temperature-corrected nor mean-normalized. Error is given per pixel in a 2048x2048 array and is saved to the second extension, `ERR'.
\end{itemize}

\noindent
Data Quality:
\begin{itemize}
    \item If the dark count rate is over the data quality threshold of 2e-3 cts/s, pixels are flagged as `hot' with a bitwise value of 16 and are otherwise given a value of 0. The data quality array is saved to the third extension, `DQ'. Flagged hot pixels are used during the dark subtraction step in the \textbf{calstis} pipeline and are propagated through to output images. Although the NUV-MAMA has few hot pixels and most are near the edge of the detector, some users doing post-processing may want to be mindful of DQ flags.
\end{itemize}

\vspace{-0.3cm}
\ssubsection{Changes in Production} \label{subsec:production_changes}

Prior to this most recent delivery, the last NUV superdark was delivered in 2009. Since then, the dark rate has evolved to a lower and more stable rate and has been characterized by a new temperature correction equation and accompanying TDCTAB. 

The previous temperature correction equation for the NUV dark files, which still applies to all pre-SM4 data, accounted solely for short-term variation in the dark count rate. The pre-SM4 temperature correction equation uses the format (\href{https://hst-docs.stsci.edu/stisdhb/chapter-4-stis-error-sources/4-1-error-sources-associated-with-pipeline-calibration-steps}{STIS DHB, 2024}; \href{https://www.stsci.edu/files/live/sites/www/files/home/hst/instrumentation/stis/documentation/instrument-science-reports/_documents/201103.pdf}{Zheng et al., 2011}):
\begin{equation} \label{eq:2}
\mbox{dark rate} = \textrm{norm} \cdot 1.805 \times 10^{20} \cdot \exp(-12211.8/\mbox{max}(T, T_{\textrm{min}}),
\end{equation}
where $T$ is the measured temperature for a given observation and $norm$ and $T_{min}$ are slowly varying functions of time, tabulated in the pre-SM4 TDCTAB, that are empirically adjusted to give a good match to the observed dark rate. The previous temperature correction equation was incorporated in an IDL script and used to create superdarks from 2009 and earlier.

To address the changes discussed in Section \ref{sec:TDCTAB}, the STIS team has developed new scripts in Python to replace the outdated IDL scripts. The new scripts incorporate the temperature scaling equation developed in 2013 (Equation \ref{eq:1}), and the newly delivered DARKFILEs are scaled by the concurrently updated TDCTAB.

\FloatBarrier

\vspace{-0.3cm}
\ssection{Analysis}\label{sec:Analysis}
\vspace{-0.3cm}
\ssubsection{Comparison to Previous Reference Files}

To validate the changes in file production, the new DARKFILEs (superdarks) were analyzed with the previous pre- and post-SM4 superdarks. Distributions of the dark count rate, flagged pixel values, and error were compared via histograms to evaluate general differences between the files and to determine if there were any anomalies in the new superdarks.

\begin{figure}[htbp]
    \centering
    \includegraphics[width=1\linewidth]{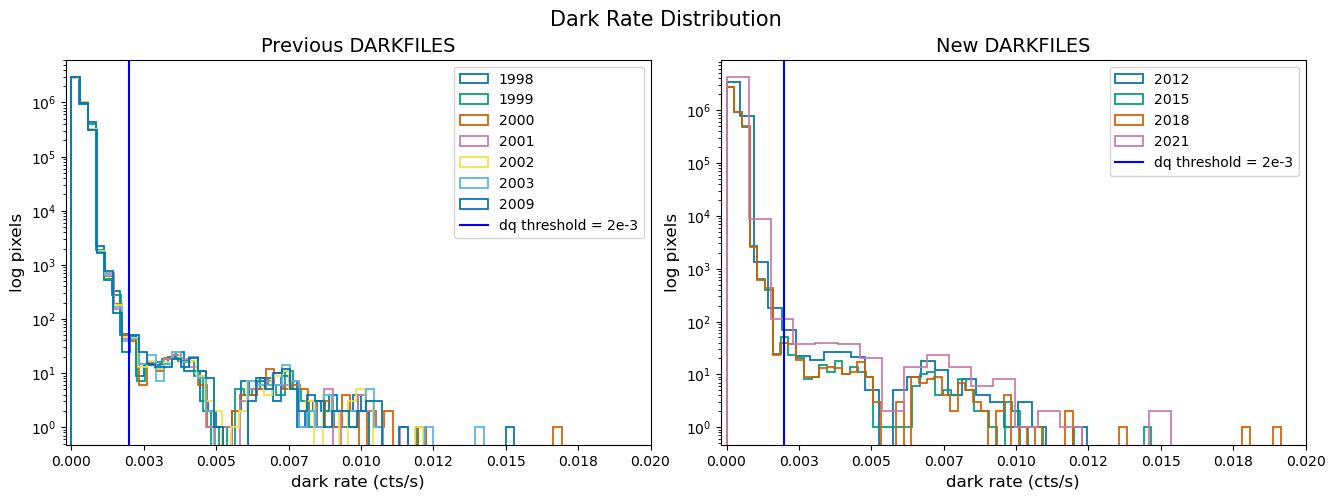}
    \caption{Histogram of the dark count rate for previous superdarks (left) and recently released superdarks (right). Both plots include the data quality (DQ) threshold, where any pixels with a value above that threshold are flagged as hot pixels and displayed in more detail in Figure \ref{fig:dq_hist}.}
    \label{fig:dark_rate_hist}
\end{figure}

No major changes to the overall count rate in the new superdarks were expected, as all are mean normalized to the same value of 2.84e-4 cts/s; however, the distribution of the dark count rate in the new superdarks was compared to the previous superdarks to check for unusual outliers and to evaluate if a change to the data quality threshold was needed. As seen in Figure \ref{fig:dark_rate_hist}, the dark count rates for the new superdarks follow a similar distribution to the previous reference files, with a slight increase in count rate on the pixels above the data quality threshold. The changes in count rate above the data quality threshold were expected, as the total number of hot pixels and their count rates increase steadily every year. Because of the increase in hot pixels, the data quality threshold was tested at different limits to evaluate if any changes were necessary, but ultimately the threshold was determined to be optimal at the value adopted by previous reference files: 2e-3 cts/s.

\begin{figure}[htbp]
    \centering
    \includegraphics[width=\linewidth]{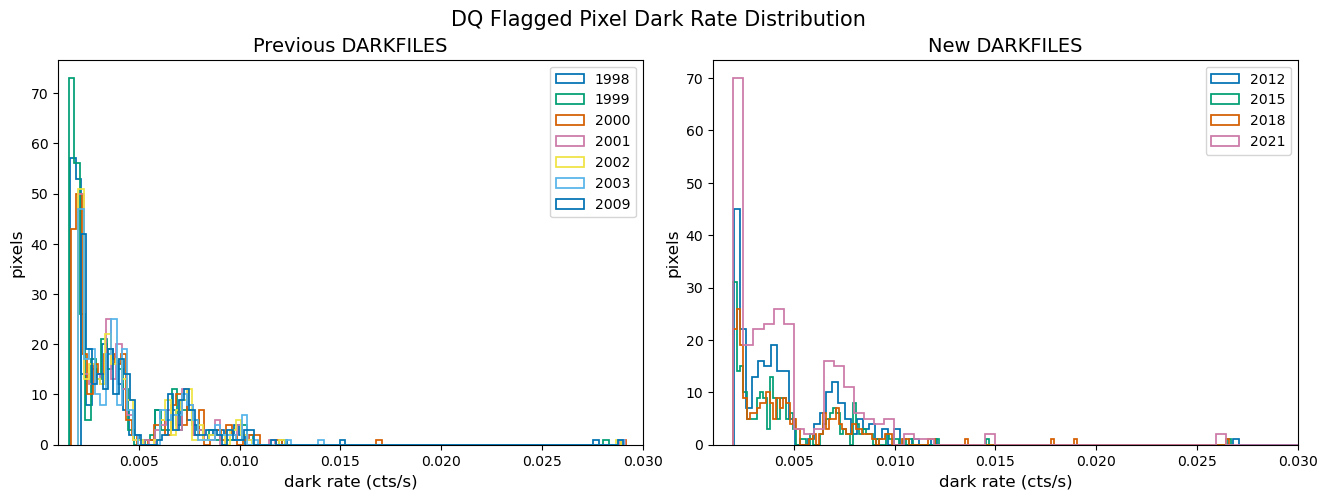}
    \caption{Histogram of the dark count rate for pixels above the data quality threshold, including previous superdarks (left) and recently released superdarks (right).}
    \label{fig:dq_hist}
\end{figure}

There are very few transient hot pixels on the NUV-MAMA detector, and because of this there is a steady increase in the number of flagged pixels from year to year (in contrast to CCDs, which usually have a large number of transient hot pixels and a variable number of flagged pixels from year to year). In Figure \ref{fig:dq_hist}, we examine the count rate distribution for the hot pixels more closely. The number of hot pixels on the NUV-MAMA detector has been increasing at an average rate of $\sim5$ pixels per year pre-SM4 and $\sim2$ pixels per year post-SM4. The average hot pixel count rate has also been increasing, with an average count rate in 2009 of 4.6e-3 cts/s (hot pixels accounted for 0.09\% of the total count rate on the detector) and an average count rate in 2021 of 5.5e-3 cts/s (hot pixels accounted for 0.12\% of the total count rate on the detector).

\FloatBarrier

\begin{figure}[htbp]
    \centering
    \includegraphics[width=\linewidth]{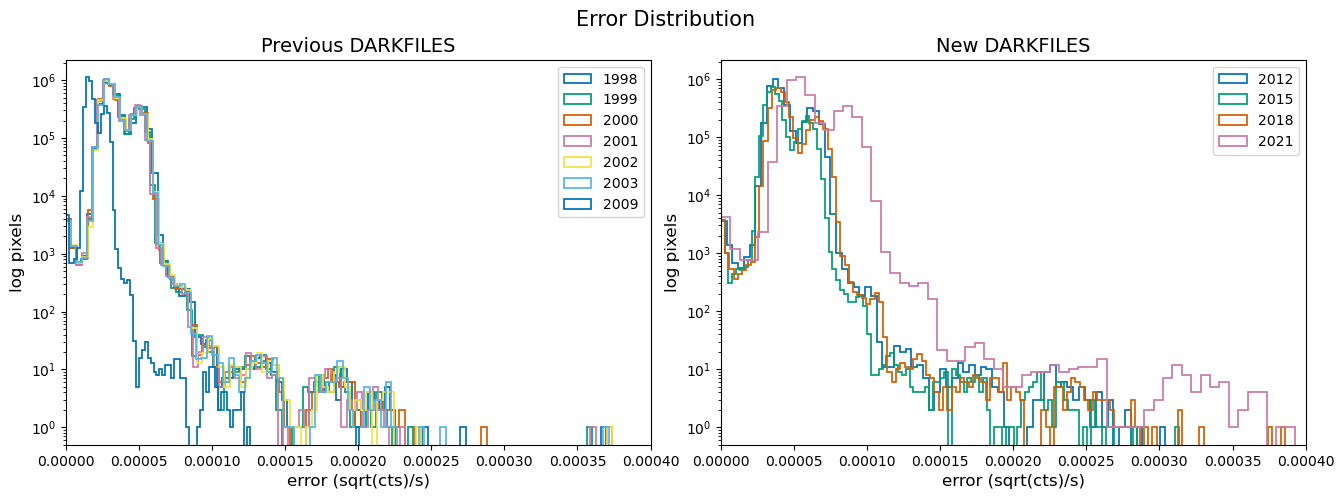}
    \caption{Histogram of the statistical error on the count rate for previous superdarks (left) and recently released superdarks (right).}
    \label{fig:error_hist}
\end{figure}

Small changes in the error distribution were expected with an increase in the number of hot pixels. To check that the error on the dark count rate for the new superdarks followed a similar distribution to that of the reference superdarks, we plotted the statistical error on the count rate in Figure \ref{fig:error_hist}. The increase in the total number and count rate of the hot pixels accounted for the increase in error in the new superdarks, especially in the 2021 superdark, which had some of the ``hottest" hot pixels nearing 0.077 cts/s. Error for the new superdarks fell within the target statistical accuracy of 2\% on the mean count rate of the detector, and did not have any measurable impact on calibrations.

\FloatBarrier

\vspace{-0.3cm}
\ssubsection{Validating Files with Calibration}

To ensure that the new superdarks improved dark calibration, they were used to recalibrate nine low SNR datasets: eight quasi-stellar objects (QSOs) and one T Tauri star (see Table \ref{tab:datasets}). Low SNR datasets were chosen for analysis as any changes to the background would be more statistically significant and easier to quantify. These datasets were chosen to span the years 2012 to 2022 to test all four new superdarks at the appropriate ``USEAFTER" date. Improvements were measured through changes to the SNR and root mean square (RMS) of regions along the extracted spectra and background. Additional improvements were measured though changes to the slope of the background subtraction calibration step compared to calibration done with the previous superdarks.

\begin{table}[htbp]
\centering
\caption{List of Low SNR Datasets}
\begin{tabular}{|l|l|l|l|}
\hline
\textbf{Observation ID} & \textbf{Type} & \textbf{Observation Start Date} \\ \hline
obrr02010               & QSO                            & 2012-05-20                      \\ \hline
oc9b02010               & QSO                            & 2014-02-26                      \\ \hline
ocod04020               & QSO                            & 2015-06-20                      \\ \hline
odsp01010               & QSO                            & 2018-09-16                      \\ \hline
odn502010               & QSO                            & 2019-04-26                      \\ \hline
oeb616010               & QSO                            & 2020-12-09                      \\ \hline
oeb611010               & QSO                           & 2021-08-21                      \\ \hline
oeb607010               & QSO                            & 2022-02-13                      \\ \hline
oeso3s010               & T Tauri Star                   & 2022-05-11                      \\ \hline
\end{tabular}
\label{tab:datasets}
\end{table}

SNR was measured in three 20 angstrom wide regions along the extracted spectra, avoiding absorption and emission lines: 2460-2480 \AA, 2540-2560 \AA, and 2990-3010 \AA. In the first region, 2460-2480 \AA, there was an average SNR increase of about 3\% when comparing calibration with the newest superdarks to the old superdark for the 9 datasets. In the second region, 2540-2560 \AA, there was an average SNR increase of about 4\%. In the third region, 2990-3010 \AA, there was an average SNR increase of about 2\%.

The RMS was calculated over both the whole extracted spectra and over the extracted background spectra. The standard background region parameters for MAMA first-order modes were used (as detailed in the \href{https://hst-docs.stsci.edu/stisdhb/chapter-3-stis-calibration/3-4-descriptions-of-calibration-steps#id-3.4DescriptionsofCalibrationSteps-3.4.2}{Chapter 3.4.2 STIS DHB, 2024}): 300 pixels below the science spectrum (or A2CENTER) with an extraction box size of 5 pixels. A decrease in RMS indicates less variation in the data and decreased noise. For the source spectra, there was an average decrease in the RMS of 2\%. The largest statistical difference was seen in the background, with an average decrease in the RMS of about 16\% in the nine datasets.

\begin{figure}[htbp]
    \centering
    \includegraphics[width=\linewidth]{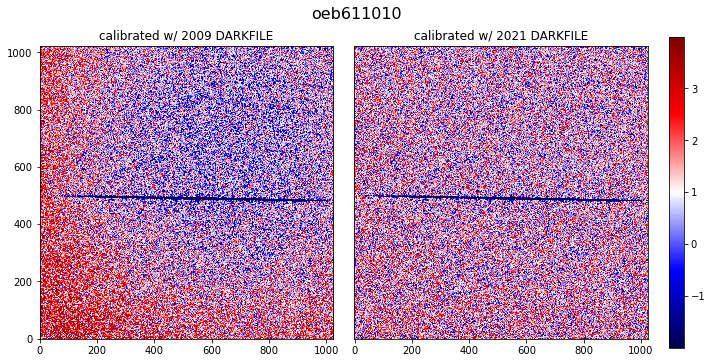}
    \caption{The FLT (background-subtracted) image of the ULLYSES dataset in which the ``ring" artifact was first identified,  reduced with the old pipeline superdark (left). When calibrated with a superdark created from contemporaneous dark exposures, the artifact is removed (right). Both images are mean-normalized and have the same scaling.}
    \label{fig:ring}
\end{figure}

\begin{figure}[htbp]
    \centering
    \includegraphics[width=\linewidth]{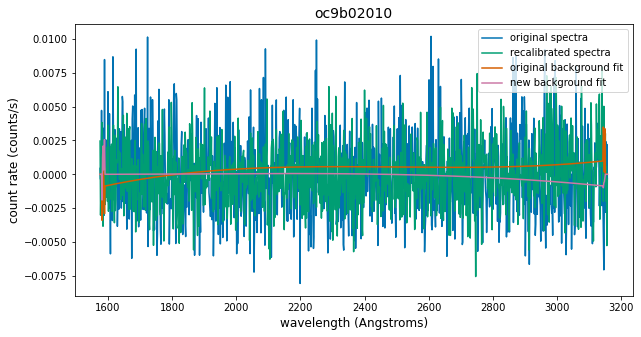}
    \caption{Comparison of extracted background spectra and fit at y=300 calibrated with the previous 2009 superdark (blue and orange) and the newer superdark (green and pink).}
    \label{fig:bkg}
\end{figure}

During the Hubble UV Legacy Library of Young Stars as Essential Standards (\href{https://ullyses.stsci.edu/}{ULLYSES}) Project, an artifact was found in the background of some very low SNR datasets, indicating updated reference files were needed for the time period after the dark count rate dropped to nominal levels, around 2012. The ring-shaped artifact could be seen in the background-corrected 2D images, corresponding to areas of higher dark rate on the NUV-MAMA detector: near the edges and especially concentrated in the bottom-left corner (different from the FUV-MAMA, where the glow region is in the upper-center of the detector). The presence of the ring artifact indicated over-subtraction of the dark current with the reference file created in 2009. When the data are calibrated with the newer superdarks, the background noise becomes more uniform, which can be seen in both the 2D image background (Figure \ref{fig:ring}) and the background fit from calibrating the spectra (Figure \ref{fig:bkg}). The ring artifact completely disappears from the image and the background fit becomes more stable with an overall reduction in noise.

\FloatBarrier

\vspace{-0.3cm}
\ssection{Summary}\label{sec:Summary}
Using data from Cycles 19 to 30, a new set of superdark reference files has been released for all supported STIS NUV observing modes. These files have been created with new code to reflect an updated temperature and time correction equation and accompanying TDCTAB that more closely models the dark current trends. In tested datasets, the new superdarks provide an average 3\% increase in SNR and smooth out a ring artifact left previously by oversubtraction of the dark rate. In February 2025, the four superdarks and the TDCTAB were delivered to the calibration pipeline and all prior archival data were reprocessed. To ensure updated darks are being used for calibration, users can check the processing date header keyword (\texttt{PROCTIME}) or the dark reference filename (\texttt{*drk.fits}) in their science data.

\vspace{-0.3cm}
\ssectionstar{Acknowledgements}
The authors thank Matthew Dallas, Daniel Stapleton, and Dan Welty for their advice and input on the creation and validation of the new superdarks.

\vspace{-0.3cm}
\ssectionstar{Change History for STIS ISR 2026-06}\label{sec:History}
\vspace{-0.3cm}
Version 1: 25 September 2026 - Original Document 
\\Version 2: 14 October 2026 - Updated Figure \ref{fig:dark_monitor}

\vspace{-0.3cm}
\ssectionstar{References}\label{sec:References}
\vspace{-0.3cm}

\noindent
Brown, J., ``STIS NUV MAMA Dark Monitor,'' \href{https://archive.stsci.edu/proposal_search.php?id=17646&mission=hst}{HST Proposal 17646}, 2024.
\\ \\
\noindent
Colin Cox, ``Dark count rates in the STIS MAMA,'' \href{https://www.stsci.edu/files/live/sites/www/files/home/hst/instrumentation/stis/documentation/instrument-science-reports/_documents/2013_01.pdf}{Instrument Science Report STIS 2013-01}, June 2013.
\\ \\
\noindent
Jones, A., ``STIS NUV MAMA Dark Monitor,'' \href{https://archive.stsci.edu/proposal_search.php?id=17391&mission=hst}{HST Proposal 17391}, 2023.
\\ \\
\noindent
Rickman, E., and Brown, J. et al., 2025, \href{https://hst-docs.stsci.edu/stisihb}{STIS Instrument Handbook}, Version 25.0, (Baltimore: STScI)
\\ \\
\noindent
Rickman, E., Brown J., et al., 2024, \href{https://hst-docs.stsci.edu/stisdhb}{STIS Data Handbook}, Version 8.0, (Baltimore: STScI)
\\ \\
\noindent
Wei Zheng, Charles Proffitt, and David Sahnow, ``Dark Rate of the STIS NUV Detector," \href{https://www.stsci.edu/files/live/sites/www/files/home/hst/instrumentation/stis/documentation/instrument-science-reports/_documents/201103.pdf}{Instrument Science Report STIS 2011-03}, September 2011.
\\ \\

\newpage
\vspace{-0.3cm}

\end{document}